\documentclass[aps,prl,twocolumn,superscriptaddress,showpacs]{revtex4}

\usepackage{graphicx}

\begin{document}

\title{Enhanced tunneling magnetoresistance in Fe$\mid$ZnSe double junctions}

\author{J. Peralta-Ramos}
\email[Corresponding author: ]{peralta@cnea.gov.ar}

\affiliation{Departamento de F\'isica, Centro At\'omico Constituyentes, Comisi\'on 
Nacional de Energ\'ia At\'omica, Buenos Aires, Argentina}

\author{A. M. Llois}
\affiliation{Departamento de F\'isica, Centro At\'omico Constituyentes, Comisi\'on 
Nacional de Energ\'ia At\'omica, Buenos Aires, Argentina}

\date{\today}

\begin{abstract}

We calculate the tunneling magnetoresistance (TMR) 
of Fe$\mid$ZnSe$\mid$Fe$\mid$ZnSe$\mid$Fe (001) double magnetic tunnel
junctions as a function of the in-between Fe layer's thickness, and compare these 
results with those of Fe$\mid$ZnSe$\mid$Fe simple junctions. The electronic band 
structures are 
modeled by a parametrized tight-binding Hamiltonian fitted to {\it ab initio} 
calculations, 
and the conductance is 
calculated within the Landauer formalism expressed in terms of Green's functions.
We find that the conductances for each spin channel and the TMR strongly 
depend on the in-between Fe layer's 
thickness, and that in some cases they are enhanced with respect to simple junctions,
in qualitative agreement with recent experimental studies performed on similar systems. 
By using a 2D
double junction as a simplified system, we show that the conductance 
enhancement can be explained in terms of the junctions energy spectrum.
These results 
are relevant for spintronics because they demonstrate that the TMR in
double junctions can be tuned and enhanced
by varying the in-between metallic layer's thickness.

\end{abstract}


\maketitle

A magnetic tunnel junction (MTJ) consists of two ferromagnetic electrodes separated by
a thin non-conducting barrier. 
It is experimentally observed that the conductance of a MTJ depends
on the relative orientation of the electrodes' magnetization, and because of this, 
during the last years a lot of 
attention has been paid to the investigation of MTJs as promising candidates for 
application in spintronic devices, such as read heads and 
magnetic random 
access memories (for reviews, see [1] and the references therein).

One of the challenges, that has to be 
overcome for 
practical applications, is to reach higher values of the tunneling magnetoresistance
ratio (TMR), defined as TMR$=[(\Gamma_{P}-\Gamma_{AP})/\Gamma_P]\times 100 \%$, 
where $\Gamma_P$ and $\Gamma_{AP}$ are the conductances measured for the parallel (P) and 
antiparallel (AP) magnetization of the electrodes. Several possibilities are now being
considered: to use highly polarized materials ({\it half-metals}) or 
diluted magnetic semiconductors as parts of MTJs, 
to produce junctions with almost perfect interfaces, 
and to use  
double magnetic tunnel junctions 
(DMTJs),
in which metallic layers are inserted inside the semiconductor barrier of a 
MTJ. In this work we explore the latter alternative, and focus our attention
on the dependence 
of the TMR on the in-between metallic layer's thickness.

Since X. Zhang {\it et al} [2] suggested to use DMTJs, several groups 
[3-6]
have theoretically shown that DMTJs exhibit richer spin-dependent transport 
properties than MTJs and that the TMR can be higher than that of MTJs,
but only 
very recently 
could these DMTJs be fabricated [7,8].
T. Nozaki {\it et al} [7] 
have recently measured the tunnel magnetoresistance of epitaxial
Fe$\mid$MgO$\mid$Fe$\mid$MgO$\mid$Fe (001) DMTJs at room 
temperature, 
and found an enhancement of the 
TMR with respect to MTJs (53 $\%$ for DMTJs versus 44 $\%$ for MTJs at low bias), 
indicating that DMTJs may 
present an advantage over simple junctions for their use in spintronics.

As far as we know, up to now  
the only theoretical studies of DMTJs {\it with magnetic layers in between the 
semiconductor} 
were made within the 
free electron model (that
cannot reproduce the
decay rates inside the semiconductor of evanescent states with different
symmetry), and using rectangular potential profiles [2,3,5,14]. 
Moreover, these studies analyzed the
dependence of TMR on the applied bias voltage and not on the in-between metallic 
layer's thickness, as we do in this work.
For this reason, in this paper transport through Fe($\infty$)$\mid$ZnSe($a$)$\mid$Fe($\infty$) 
(001) MTJs and 
through Fe($\infty$)$\mid$ZnSe($b$)$\mid$Fe($c$)$\mid$ZnSe($b$)$\mid$Fe($\infty$) (001) 
DMTJs 
is theoretically investigated using a realistic tight-binding (TB) Hamiltonian
to obtain the electronic structure of the junctions. Fe($\infty$) are semi-infinite 
electrodes, and
$a$, $b$ and $c$ denote thicknesses.
The systems studied are epitaxial, and we restrict to zero temperature, infinitesimal 
bias 
voltage and elastic transport. We choose Fe$\mid$ZnSe 
because it can be grown epitaxially and there is very 
little interdiffusion at the interfaces, thus producing crystalline 
junctions in which there are no magnetically dead Fe layers [9,10]. Moreover,
in contrast to what happens in
Fe$\mid$MgO based junctions,
there is no oxidation of the interfacial Fe layers, which is known to be 
detrimental to TMR [11]. To obtain a clearer insight into the physics involved
in transport through double tunnel junctions, we also calculate the conductance
through a simplified {\it two-dimensional} tunnel junction (2DDJ).

The conductances are calculated from the active region's 
Green's function 
$G_S^\sigma=[\hat{1}E-H_S^\sigma-\Sigma_L^\sigma-\Sigma_R^\sigma]^{-1}$, 
where $\hat{1}$ stands 
for the unit matrix, 
$H_S^\sigma$ is the Hamiltonian corresponding to the active region, 
$\Sigma_{L/R}^\sigma$ are the self-energies 
describing the interaction of the active region with the left (L) or right (R) 
electrodes ($\sigma$ corresponds to the majority or minority spin
channels), and 'active region' stands for whatever is sandwiched by the 
electrodes. For DMTJs, the active region consists of an 'in-between metal region' 
(IBMR) sandwiched by {\it two identical} 'semiconductor regions' (SCR), while for 
MTJs the
active region is simply the SCR. The energy $E$ is actually $E_F+i\eta$, $E_F$ being
the Fermi level of the system, and we take 
$\eta \rightarrow 0^+$.
The self-energies are given by 
$\Sigma_L^\sigma=H_{LS}^{\dagger} g_L^\sigma H_{LS}$ and $\Sigma_R^\sigma=
H_{RS}^{\dagger} g_R^\sigma 
H_{RS}$, where $H_{LS}$ and $H_{RS}$ describe the coupling of the active
region with 
the electrodes, and $g_{L/R}^\sigma$ are the surface Green's functions for each 
electrode. These surface Green's functions are calculated using a semi-analytical 
method [12] and are exact within our TB approximation. The 
transmission probability $T^\sigma$ is given by [13]
$T^\sigma(k_{//},E_F)=Tr~ [\Delta_L^\sigma G_S^\sigma \Delta_R^\sigma 
G_S^{\sigma \dagger}]$
where $\Delta_{L/R}^\sigma=i (\Sigma_{L/R}^\sigma-\Sigma_{L/R}^{\sigma \dagger})$, 
while the conductance is given by 
\begin{equation}
\Gamma^\sigma(E_F)=\frac{e^2}{h}\frac{1}{N_{k_{//}}} \sum_{k_{//}} T^\sigma (k_{//},E_F)
\end{equation}
where $N_{k_{//}}$ is the total number of wave vectors parallel to the interface 
that we consider (in our case 5000 is enough to achieve convergence in 
$\Gamma$). 

We start our discussion with the 2DDJs case, which are of the type
M($\infty$)$\mid$S$\mid$M$\mid$S$\mid$M($\infty$), where M($\infty$) are 
semi-infinite 
{\it paramagnetic} metallic electrodes, 
S is a 
semiconductor
and M is a metal (the same as the electrodes). 
The metal and semiconductor have the 
same structure, a {\it square} Bravais lattice with two atoms per unit cell, and are
periodic in the direction perpendicular to the transport direction. The 2DDJs electronic 
structure
is modeled by a 2nd nearest neighbors TB Hamiltonian with one $s$ orbital per atom. 
The TB parameters are chosen to make $E_F$ fall in the middle of the semiconductor's
band gap (of 0.5 eV).
The SC and IBM regions are varied between 3.2 \AA~ and 32 \AA~. 

It is found that for certain
thicknesses of the IBMR the conductance
presents peaks in which it is enhanced by 1 to 4 orders of magnitude, 
as can be seen in 
Fig. 1 for a 2DDJ with a SCR of 12.8 \AA~.  
This effect can be explained in 
terms of the
active region's density of states (DOS), obtained from its Green's function
$G_S$. When the conductance is enhanced, partial
density of states (PDOS) calculations indicate that there exist 
states at $E_F$ extended throughout the whole junction, so in that
situations transport occurs through resonant states. When this happens, the 
2DDJs conductances are higher than the corresponding ones of simple 2D junctions. 
These 
results 
are consistent with those of Z. Zheng and coworkers for a DMTJ with a non-magnetic 
in-between metal [14].
\begin{figure}[h]
\scalebox{0.5}{\includegraphics{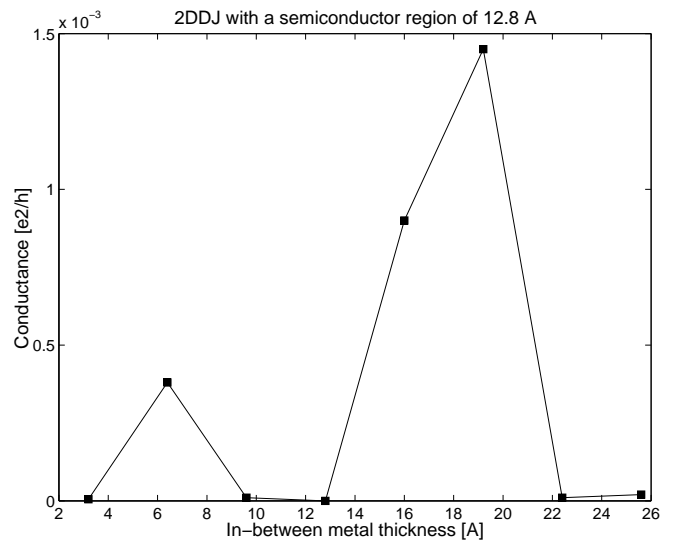}}
\caption{Conductance as a function of in-between metal thickness for 2DDJs with
a semiconductor region of 12.8 \AA~.}
\end{figure}

Fig. 2 shows the maximum ratio between
the conductance of 2DDJs and 2D simple junctions, as a function of
the SCR thickness. The maximum attainable ratio is of 146 $\%$ and occurs for a
SCR thickness of 9.6 \AA~ and an IBMR thickness of 19.2 \AA~. For thinner SCRs
the ratio is nearly constant and roughly 140 $\%$, but beyond 12.8 \AA~ the 
enhancement effect is lost. 
Having mentioned
the main results for the 2DDJs, we go on to discuss the details for the 
three-dimensional Fe$\mid$ZnSe DMTJs.
\begin{figure}[h]
\scalebox{0.5}{\includegraphics{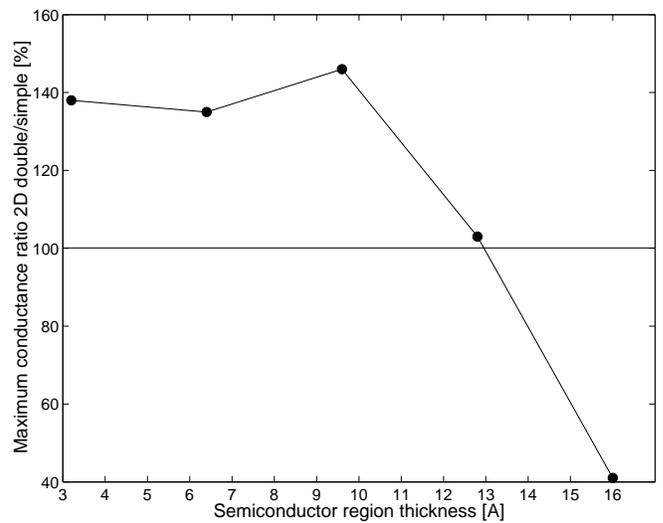}}
\caption{Maximum ratio between the conductances of 2DDJs and simple 2D junctions 
as a function of 
the semiconductor region thickness.}
\end{figure}

Fig. 3 shows schematically the structure of simple and double junctions, which are
periodic in the {\it x-y} plane, and the different magnetic configurations considered,
parallel (P) and antiparallel (AP). Since the coercive field of the
electrode and the in-between Fe layers is different, the magnetic configurations shown are
experimentally attainable [7]. 
\begin{figure}[h]
\scalebox{0.3}{\includegraphics{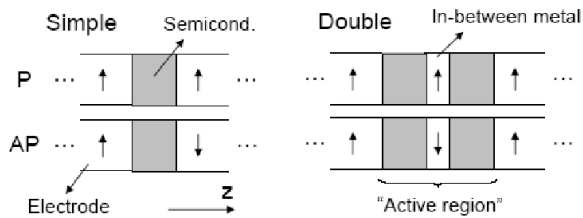}}
\vspace{-0.8cm}
\caption{Schematic structure of simple and double junctions, and magnetic 
configurations considered: parallel (P) and antiparallel (AP). The junctions are
periodic in the {\it x-y} plane, and the electrodes are semi-infinite. The arrows 
indicate the magnetization direction of the metallic regions.}
\end{figure}

Fig. 4 shows the structure of a simple Fe$\mid$ZnSe junction with a SCR of
5.67 \AA~, along the $z$ 
direction (which is the direction of transport). The
BCC Fe lattice parameter is $2.87$ \AA~, and that of zincblende ZnSe is $5.67$ \AA~.
\begin{figure}[h]
\scalebox{0.35}{\includegraphics{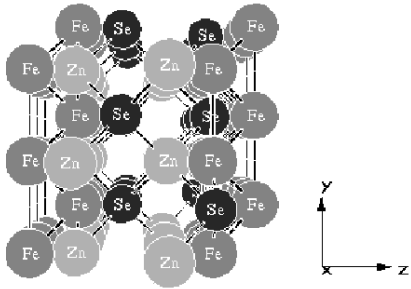}}
\vspace{-0.4cm}
\caption{Interfacial structure of a simple junction of BCC Fe electrodes and zincblende ZnSe
semiconducting spacer along the (001) direction, which is the transport direction.
In the example shown, the semiconductor's thickness is of 
5.67 \AA~.}
\end{figure}

The electronic structure of the junctions is modeled by a parametrized  
2nd nearest neighbors {\it spd} 
TB Hamiltonian fitted to {\it ab initio} calculations [15,16], in which
the hoppings between the Fe atoms and the (Zn,Se) atoms are calculated using
Shiba's rules and Andersen's scaling law 
[17]. The Fe $d$ bands are spin split by $\mu J_{dd}$, where $\mu=2.2 ~\mu_B$ 
is the
experimental magnetic moment of Fe and $J_{dd}=1.16$ eV is the exchange integral 
between
$d$ orbitals ($\mu_B$ is Bohr's magneton). With these values for $\mu$ and $J_{dd}$, the  
Fe $d$ bands spin spitting
is very well reproduced [15].
The ZnSe band 
structure is rigidly shifted to make the iron Fermi energy fall 1 eV above 
the ZnSe valence band and 1.1 eV below
the conduction band, as indicated by photoemission experiments [9].

For simple junctions, we find that the conductances decay almost exponentially
with semiconductor thickness, and that the TMR increases and is
always positive (or direct), reaching a value of
90 $\%$ for a semiconductor thickness 
of 34 \AA~. Our results are in very good agreement with
the {\it ab initio} results of MacLaren and coworkers [18]. 

For double junctions, we vary
the SCR thickness between 5.67 \AA~ and 28.35 \AA~, and the IBMR thickness
between 2.87 \AA~ and 22.96 \AA~. We find that the
TMR and conductances strongly depend on the in-between metal thickness, and that 
for certain thickness
combinations of the 
SC and IBM regions they can be higher than those corresponding to a MTJ, in agreement
with the results of L. Sheng and coworkers [3].
The maximum ratio of DMTJs to simple MTJs conductances obtained is of 322 $\%$, 
and occurs for the P majority channel corresponding to
SC regions of 22.7 \AA~ and an IBM region of 21 \AA~. This large conductance ratio, 
which is pointing toward the existence of resonant states (confirmed by our DOS 
calculations), does not mean
that the DMTJs TMR is going to be much larger than the MTJs one, although
in general it is. In this 
particular case, the TMR value 
is of 97.9 $\%$ (the corresponding MTJ's value is 
63.8 $\%$), but in other cases the TMR values are greatly enhanced {\it even in the 
absence of resonances}. We find that the TMR enhancement can be a result
of: (i) a drop in the 
conductance of some spin channels, while the conductances of other channels 
remain of the same order of magnitude
as those in MTJs, or (ii) an increase in the conductance of one
particular spin channel due to resonant tunneling. Both effects are produced by a change
in the active region's DOS near $E_F$, induced by the presence of the in-between
Fe layers.

As an example of resonance conductance enhancement, 
we show in Fig. 5 the active region's total DOS 
at $E_F$,
as a 
function of the IBMR thickness and for a DMTJ with a SCR of
22.7 \AA~. An increase in one order of magnitude appears at an IBMR of 2.87 \AA~ for 
the P majority channel and 
for the AP minority channel, and a smaller increase appears at an IBMR of 
8.6 \AA~ for the AP majority channel, 
while for the
other cases the DOS is almost constant. These peaks coincide with a conductance
enhancement in these three channels, as it can be seen in the lower panel of Fig. 5,
indicating that the origin of the conductance enhancement is the same
as in 2DDJs, namely resonant tunneling.
\begin{figure}[h]
\scalebox{0.5}{\includegraphics{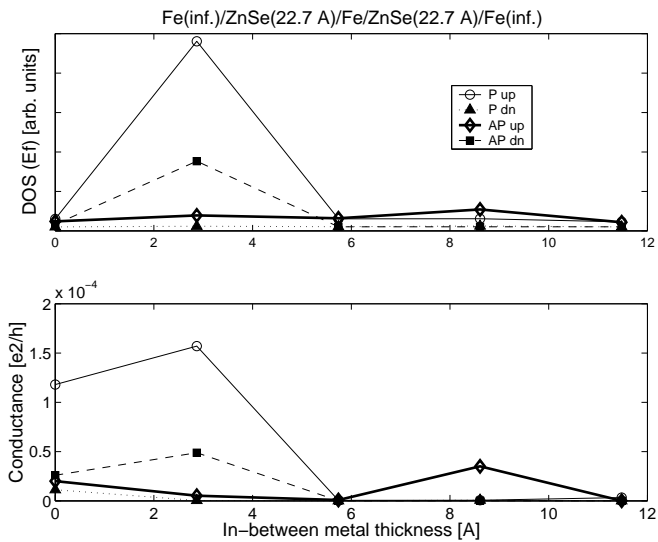}}
\caption{Active region's density of states and conductances at $E_F$ as a 
function of in-between metal thickness, for
a semiconductor region of 22.7 \AA~.}
\end{figure}

To visualize the interplay among the conductance values of the different channels
and configurations, Fig. 6 shows the conductances and TMR values
for a given
DMTJ with a SCR of 17 \AA~ 
and those of the corresponding simple MTJ,  
as a function of the IBMR thickness. It is seen that, already for 6 \AA~ 
of Fe, the TMR is 1.5 times higher than that of a simple MTJ, although 
the conductances are, 
in general, a little
bit smaller. This is similar to what happens 
in all the cases studied.
It is noticeable that for very 
thin Fe layers the TMR obtained for this DMTJ is negative, and that for IBMR 
thicknesses
in the range 12-23 \AA~ the TMR is almost constant. This also
happens for the other SCR thicknesses studied, and it is different to the 
damped oscillatory
behavior that it is obtained using rectangular potential profiles and a non-magnetic
in-between metal [4].
\begin{figure}[h]
\scalebox{0.5}{\includegraphics{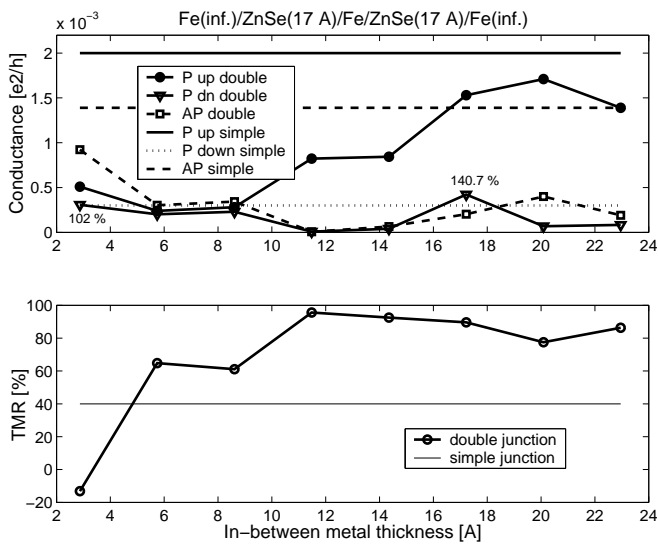}}
\caption{Conductance and TMR values of a Fe$\mid$ZnSe double magnetic junction with 
two semiconductor 
regions of 17 \AA~, and those of the corresponding simple junction, 
as a function of the in-between metal thickness.}
\end{figure}

For a given SC region thickness, we look for the maximum attainable
TMR value by sweeping over
IBMR's thicknesses.
We find that for ZnSe regions
with thicknesses below 20 \AA~, the DMTJs' TMR values are 3 times higher than those 
of a simple
junction, {\it while the conductances of some spin channels remain of the same order of 
magnitude}. 
Beyond this 
thickness, 
the DMTJs TMR can be 50 times higher but 
negative (inverse TMR), although in this case the conductances {\it of all spin 
channels} 
are 4 to 6 orders of 
magnitude
smaller, 
and thus very hard to measure. There is one particular case in which this does not happen. 
For a DMTJ with a SCR of 22.7 \AA~ and an IBMR of 8.6 \AA~, we obtain a drop in the 
conductances of the P and AP minority
channels and the P majority channel, and an enhancement of 175 $\%$ in the AP majority channel 
with respect to the corresponding MTJ, which results in a negative TMR 
enhancement by a factor of -40. 

In summary, we have investigated Fe$\mid$ZnSe double magnetic tunnel junctions 
within a 
realistic
Hamiltonian model and found that the TMR values can be much higher than those
of simple junctions. 
We should mention that temperature effects, interfacial roughness, and the presence 
of defects in the DMTJs active
region may decrease the TMR values obtained in our calculations, but we believe
that our results remain qualitatively valid.
We conclude that the thickness of the in-between Fe layers in Fe$\mid$ZnSe 
DMTJs 
is an interesting degree
of freedom, which may make it possible to tune and enhance the TMR of 
these systems, making them suitable for building future spintronic devices. 
To improve our understanding
of these scarcely studied double junctions, it is highly desirable the experimental 
measurement of the TMR as 
a function of the in-between metallic layers thickness.

We are grateful to Juli\'an Milano for useful discussions. 
This work was partially funded by UBACyT-X115, 
Fundaci\'on Antorchas and PICT 03-10698. Ana Mar\'ia Llois belongs to CONICET 
(Argentina).

\vspace{0.1cm}
\begin{small}
\noindent
$[1]$ X-G. Zhang and W. H. Butler, J. Phys.: Condens. Matter {\bfseries 15}, 1603
(2003); E. Y. Tsymbal, O. N. Mryasov, and P. R. LeClair 
{\it ibid.} {\bfseries 15}, 109 (2003)\\ 
$[2]$ X. Zhang, B-Z Li, G. Sun, and F-C. Pu, Phys. Rev. B 
{\bfseries 56}, 5484 (1997)\\
$[3]$ L. Sheng, Y. Chen, H. Y. Teng, and C. S. Ting, Phys. Rev. B {\bfseries 59},
480 (1999)\\
$[4]$ M. Chshiev, D. Stoeffler, A. Vedyayev, and K. Ounadjela, Europhys. Lett. 
{\bfseries 58}, 257 (2002)\\
$[5]$ B. Wang, Y. Guo, and B-L. Gu, J. Appl. Phys. {\bfseries 91}, 1318 (2002)\\
$[6]$ F. Giazotto, Fabio Taddei, Rosario Fazio, and Fabio Beltram, Appl. Phys. Lett. 
{\bfseries 82}, 2449 (2003)\\ 
$[7]$ T. Nozaki {\it et al}, Appl. Phys. Lett. {\bfseries 86}, 082501 (2005)\\
$[8]$ Z. Zeng {\it et al}, Phys. Rev. B {\bfseries 72}, 054419 (2005); 
J. H. Lee {\it et al}, J. Magn. Magn. Mater. {\bfseries 286}, 
138 (2005)\\
$[9]$ M. Eddrief {\it et al}, Appl. Phys. Lett. {\bfseries 81}, 4553 (2002)\\
$[10]$ M. Marangolo {\it et al}, Phys. Rev. Lett. {\bfseries 88}, 
217202-1 (2002)\\
$[11]$ X-G. Zhang, W. H. Butler, and A. Bandyopadhyay, 
Phys. Rev. B {\bfseries 68}, 092402 (2003)\\
$[12]$ S. Sanvito {\it et al}, Phys. Rev. B {\bfseries 59},
11936 (1999)\\
$[13]$ S. Datta, 'Electronic transport in mesoscopic systems' 
(Cambridge University Press, United Kingdom, 1999)\\
$[14]$ Z. Zheng, Y. Qi, D. Y. Xing, and J. Dong, Phys. Rev. B {\bfseries 59}, 14505 
(1999)\\
$[15]$ D. A. Papaconstantopoulos, 'Handbook of the band structure of 
elemental solids' (Plenum Press, New York, 1986)\\
$[16]$ R. Viswanatha, S. Sapra, T. Saha-Dasgupta and D. D. Sarma, 
cond-mat 0505451 v1, 18 May 2005\\
$[17]$ O. K. Andersen, Physica B {\bfseries 91}, 317 (1977)\\
$[18]$ J. M. MacLaren, X. G. Zhang, W. H. Butler and X. Wang,
Phys. Rev. B {\bfseries 59}, 5470 (1999)\\

\end{small}

\end{document}